\documentclass[journal=jacsat,manuscript=article]{achemso}

\usepackage[version=3]{mhchem} 
\usepackage{natbib}
\usepackage{graphicx}
\usepackage{amsmath}
\usepackage{multirow,makecell}
\usepackage{bm}
\usepackage{float}
\usepackage{soul}
\usepackage{xcolor}
\usepackage{subcaption}
\usepackage{mciteplus}
\usepackage{siunitx}
\usepackage{adjustbox}
\usepackage{threeparttable}
\usepackage{caption}
\usepackage{indentfirst}
\usepackage{textcomp}

\usepackage{xr-hyper}
\usepackage[colorlinks,allcolors=blue]{hyperref}
\usepackage[nameinlink,noabbrev,capitalize]{cleveref}
\SectionNumbersOn

\author{Xiaoyu Wang}
\affiliation{Chemical Sciences and Engineering Division, Argonne National Laboratory, 9700 S Cass Ave, Lemont, IL 60439}
\email{xiaoyu.wang@anl.gov}
\author{Srikanth Nayak}
\affiliation{Chemical Sciences and Engineering Division, Argonne National Laboratory, 9700 S Cass Ave, Lemont, IL 60439}
\author{Richard E. Wilson}
\affiliation{Chemical Sciences and Engineering Division, Argonne National Laboratory, 9700 S Cass Ave, Lemont, IL 60439}
\author{L. Soderholm}
\affiliation{Chemical Sciences and Engineering Division, Argonne National Laboratory, 9700 S Cass Ave, Lemont, IL 60439}
\author{Michael J. Servis}
\affiliation{Chemical Sciences and Engineering Division, Argonne National Laboratory, 9700 S Cass Ave, Lemont, IL 60439}
\email{mservis@anl.gov}

\title
  {Solvent Effects on Extractant Conformational Energetics in Liquid-Liquid Extraction: A Simulation Study of Molecular Solvents and Ionic Liquids}

\abbreviations{Uranyl,CMPO,Ionic Liquid,Molecular Dynamics,IR}

\begin{document}

\begin{tocentry}
\centering
\includegraphics[width=1.8in,
  height=1.8in,
  keepaspectratio]{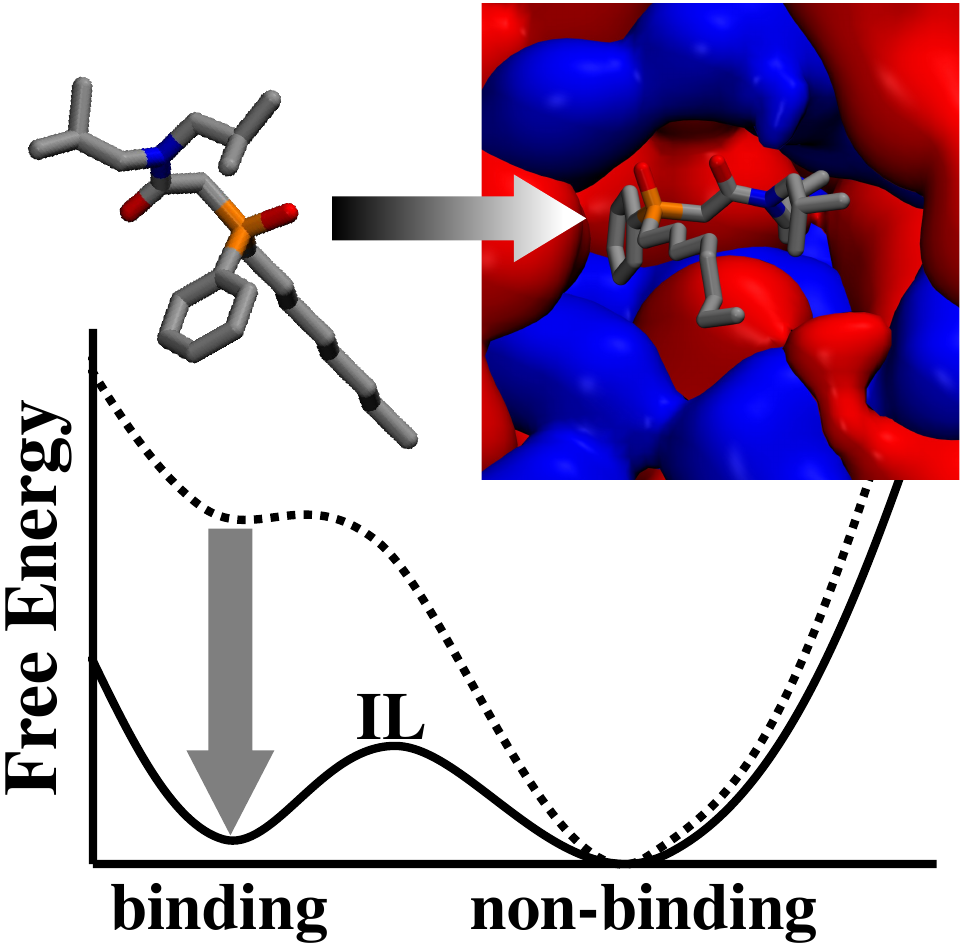}

\end{tocentry}


\newpage
\begin{abstract}

Extractant design in liquid-liquid extraction (LLE) is a research frontier of metal ion separations that typically focuses on the direct extractant-metal interactions.
However, a more detailed understanding of energetic drivers of separations beyond primary metal coordination is often lacking, including the role of solvent in the extractant phase.
In this work, we propose a new mechanism for enhancing metal-complexant energetics with nanostructured solvents.
Using molecular dynamics simulations with umbrella sampling, we find that the organic solvent can reshape the energetics of the extractant's intramolecular conformational landscape.
We calculate free energy profiles of different conformations of a representative bidentate extractant, n-octyl(phenyl)-N,N-diisobutyl carbamoyl methyl phosphinoxide (CMPO), in four different solvents: dodecane, tributyl phosphate (TBP), and dry and wet ionic liquid (IL) 1-ethyl-3-methylimidazolium bis(trifluoromethylsulfonyl)imide (\ce{[EMIM][Tf_2N]}).
By promoting reorganization of the extractant molecule into its binding conformation, our findings reveal how particular solvents can ameliorate this unfavorable step of the metal separation process.
In particular, the charge alternating nanodomains formed in ILs substantially reduce the free energy penalty associated with extractant reorganization.
Importantly, using alchemical free energy calculations, we find that this stabilization persists even when we explicitly include the extracted cation.
These findings provide insight into the energic drivers of metal ion separations and potentially suggest a new approach to designing effective separations using a molecular-level understanding of solvent effects.

\end{abstract}
\newpage


\section{Introduction}

Liquid-liquid extraction (LLE) leverages the relative solubilities of chemical species between two immiscible phases, usually an aqueous and an organic phase, to selectively recover a wide range of critical materials, including rare earths, actinides, and platinum group elements. \cite{sholl2016seven,sun2012ionic,de2021extraction}
In hydrometallurgical processes, the extractant, an amphiphilic metal complexant, selectively binds the targeted ions and solubilizes them in the nonpolar organic phase.
LLE is a free energy driven process, relying on small differences in metal ion solvation free energies between phases to generate efficient yet reversible separations.
For many separation processes, these differences are often only a few kJ/mol: many large energetic factors contribute to the extraction and cancel out, resulting in a total free energy of extraction that can be much smaller than each contribution.
For example, enthalpic metal ion binding with the complexant generally favors phase transfer, whereas extractant-solvent interactions, extractant conformational reorganization\cite{lumetta2002deliberate} and self-assembly\cite{spadina2021molecular} into molecular complexes inhibit the process.

Approaches to rational process design are typically simplified by focusing on the most significant enthalpic interactions.
For example, recent studies have focused on improving the chelating functionality of the complexant for targeted metal recovery. \cite{lumetta2002deliberate,lewis2011highly,ansari2012chemistry,mccann2016computer,ivanov2016computational,brigham2017trefoil,morgenstern2020computer,bessen2021complexation,liu2022advancing,driscoll2022noncoordinating,zhang2022advancing}
Such works typically focus on understanding and improving local metal coordination binding and extractant conformational energetics.
Although not all interactions are explicitly and properly considered when following this approach, the role of organic phase hierarchical structuring has received limited recent attention.\cite{yamamuro2011hierarchical,baldwin2018outer,servis2020hierarchical,wang2020microstructural}
In general, complexant-solvent interactions are accounted for with simple continuum models, despite the well-documented solvent effects on metal separation energetics. \cite{ansari2012chemistry,sprakel2019improving,spadina2021molecular,guilbaud2015depletion,li2019enhancing}
Herein we discuss one such structural factor---multidentate-complexant conformation stability---and its potential impact on the energetics driving a separation.

Multidentate amphiphilic molecules are a common class of extractants that utilize chelation effects to enhance the energetics of separation and to improve the selectivity of metal ions. \cite{hay2001structural,boehme2002carbamoylphosphine,zhang2022advancing,ellis2017straining}
However, for many multidentate extractants, the binding conformation is not the most favorable for the free, unbound molecule. \cite{hay2001structural}
Generally, multidentate complexants with chelating oxygen atoms bind metal ions with oxygen atoms aligned in the \textit{cis} conformation.
In contrast, free (unbound) extractant molecules typically favor the \textit{trans} conformation, where the repulsion between electronegative oxygen atoms is minimized.
For example, \citeauthor{kapoor2005di} studied single-crystal structures of unbound CMPO, as well as its chelate complex with uranyl nitrate, and found that the CMPO in the urnayl chelate complex exhibits a \textit{cis} conformation, while the unbound CMPO remains at a \textit{trans} conformation. \cite{kapoor2005di}
To quantify the energetic differences between the two molecular conformations---\textit{cis} and \textit{trans}---we consider a representative bidentate extractant with well-characterized separation behavior, CMPO.
The molecular structure and the \textit{cis}, \textit{gauche} and \textit{trans} conformations are shown in \cref{fig:cmpo}.
Our umbrella sampling MD simulations in dodecane (see below) find that the free energy difference between \textit{cis} and \textit{trans} CMPO is significant (up to 30 kJ/mol) compared to the overall extraction free energy. Considering the available thermal energy at room temperature is around 2.5kJ/mol, \textit{cis} conformation only takes up 0.0006 \% of the population.
Such a high energy barrier strongly suggests that, though the \textit{cis} conformation is relevant while bidentate chelating the metal ion, the \textit{trans} conformation dominates for the free complexant in typical nonpolar organic solvents such as dodecane. \cite{chaumont2004solvation,chaumont2006solvation}

\begin{figure}[h]
\centering
\includegraphics[width=1.0\linewidth]{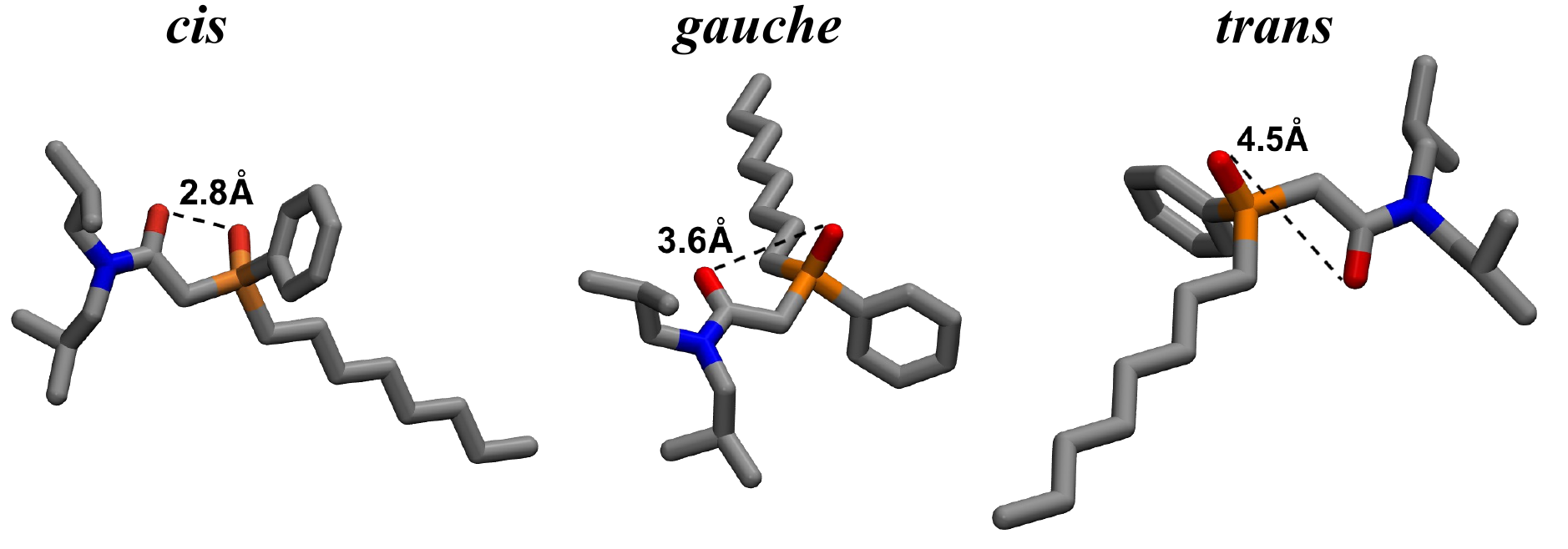}
\caption{The \textit{cis}, \textit{gauche} and \textit{trans} conformations of CMPO are shown, with corresponding oxygen-oxygen distances. Color scheme: red – oxygen, blue – nitrogen, gray – carbon, orange – phosphorus.}
\label{fig:cmpo}
\end{figure}

\begin{figure}[h]
\centering
\includegraphics[width=1.0\linewidth]{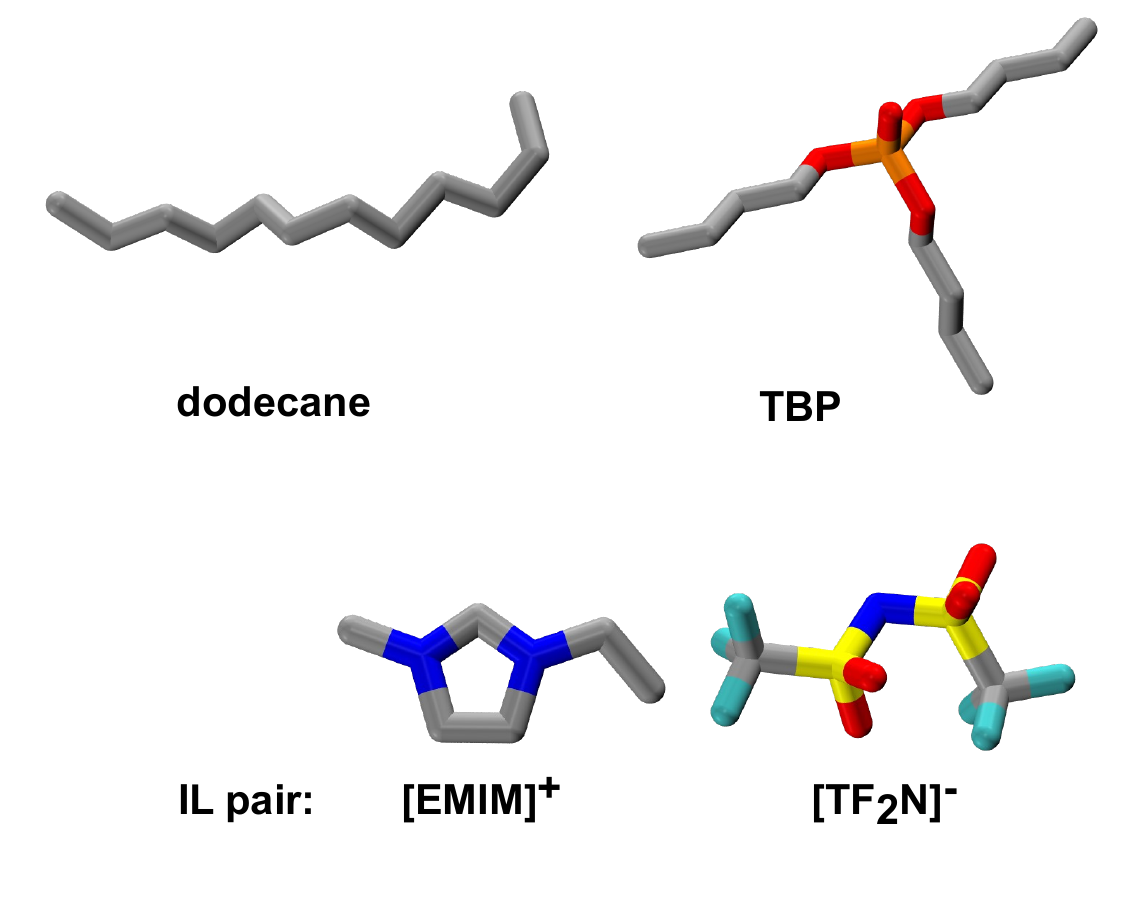}
\caption{Molecular structures of dodecane, TBP and IL pair are shown. Color scheme: red – oxygen, blue – nitrogen, gray – carbon, orange – phosphorus, cyan - fluorine.}
\label{fig:c12_tbp_IL}
\end{figure}

One ligand design strategy used to avoid the energetic cost of the extractant's conformation is to fix the extractant in the chelating conformation by adding a rigid backbone \cite{lewis2011highly}, which comes at the cost of flexibility in the binding pocket and leads to more complicated synthetic routes for the extractant molecule.
Here, we suggest that the energetic cost of such conformational reorganization can be reduced instead by tuning solvent-extractant interactions.
To broaden our understanding of the potential role of the organic solvent, we first consider two conventional solvents, dodecane and TBP (molecular structures are shown in \cref{fig:c12_tbp_IL}), specifically chosen to cover a range of polarities, and in addition they are well-characterized and commonly employed with CMPO in LLE processes. \cite{ellis2012mesoscopic}
Furthermore, we consider two model IL systems -- dry and ``wet'' (water saturated) \ce{[EMIM][Tf_2N]} (molecular structures are shown in \cref{fig:c12_tbp_IL}).
ILs are a class of solvents that have received substantial attention in recent years due in part to their unique liquid nanostructure:\cite{canongia2006nanostructural,hayes2015structure,yan2019ionic,lyu2020review} they form polar/apolar and cationic/anionic domains, \cite{canongia2006nanostructural,kashyap2013structure,hayes2015structure,wang2020microstructural} resulting in drastically different nanostructuring compared to traditional organic molecular solvents.
ILs have a broad appeal, finding their use in a wide range of applications through highly customizable organic cations and anions, which enable tailoring of physicochemical properties based on specific application needs.
They also serve as low vapor pressure replacements for the volatile organic solvents typically employed in separations.\cite{dietz2008anion,billard2013ionic}
ILs have also shown potential to enhance metal ion separations. \cite{visser2003room,dai1999solvent,kubota2011application,rout2009extraction,shen2011extraction,turanov2020solvent} 
For example, with CMPO, the distribution ratio for uranyl extraction increases from $\sim$1 to $\sim$20 (for an initial aqueous of 0.01 M \ce{HNO3}) when the extractant phase solvent is changed from dodecane to IL. \cite{visser2003room,visser2003uranyl}
The increase in this case is due, at least in part, to the different extraction mechanisms enabled by the IL, such as cation exchange using the IL cation, which makes direct comparison to conventional solvents challenging.
Here, we compare solvents in a manner that is agnostic to the particular extraction mechanism.

In the following, we present MD simulations with a free energy estimation technique, umbrella sampling, to show how, compared to typical solvents such as dodecane, TBP, and, to a greater extent, IL, can significantly bias the conformational free energy profile of the extractant, stabilizing the metal-binding \textit{cis}-CMPO conformation.
To verify that this effect persists even when the extractant binds with the metal ion, we also carry out IL-phase solvation free-energy calculations in which an uranyl ion is ``annihilated'' while bound to both \textit{cis}-restrained and free CMPO.
The ``annihilation'' in alchemical free energy calculations is achieved by turning off in a stepwise fashion pairwise interactions so that the particle of interest is decoupled from the system. This approach provides a quantitative calculation of the corresponding solvation free energy of that particle.
By comparing the two end states, restrained \textit{cis} and free CMPO, we are able to separate the stabilization effects contributed by the metal-extractant complexation and the IL-phase solvent effect, finding that the solvent effect still contributes substantially to favoring the binding conformation.
Overall, in contrast to traditional research approaches focused on designing task-specific extractants by tailoring their molecular structures and improving their functionalities, our findings suggest how structure-directing solvents could be used to control extractant conformational energetics, thereby suggesting a new route for enhancing LLE separation efficiency. 

\section{Experiments}

\subsection{Materials}

Caution! U-238 is an alpha-emitting isotope.
All experiments described were performed in specially designed laboratories with negative pressure fume hoods and gloveboxes, using strict radiological controls.

We took CMPO from Argonne stock, which was purified by the methodology in \citeauthor{gatrone1987synthesis}'s work \cite{gatrone1987synthesis}.
The reported purity for this synthesis and purification process is $\geq$99\%.
To confirm purity, we also performed $^{31}$P-NMR on the CMPO, finding no contaminants in detectable quantities. We put the NMR data into the SI. The \ce{[EMIM][Tf_2N]} ($>$98\%) were obtained from Sigma-Aldrich. 
NaNO$_3$ ($>$99\%) and HNO$_3$ (67-70\%) were obtained from Fisher Chemicals.
UO$_2$(NO$_3$)$_2$.6H$_2$O was obtained from the Argonne National Laboratory stock.
All the chemicals were used as received.

\subsection{FT-IR Spectra by Density Functional Theory (DFT) Calculations}

In the DFT calculations, we use a complex consisting of one uranyl, with one bidentate CMPO and two bidentate nitrate anions to fill the equatorial coordination vacancies and neutralize the complex.
The geometry of the uranyl-CMPO complex is first optimized with the hybrid functional theory M06-2X \cite{zhao2008m06} in Gaussian 16 \cite{g16}; the Stuttgart/Dresden relativistic large core potential (RLC) basis set (ECP), obtained from Basis Set Exchange \cite{pritchard2019new}, is used for uranium, while the basis set 6-311++G(d,p) is used for other elements in the complex.
Single-point energy at the same level of theory as the optimization step with frequency calculation is performed.
The stretching frequencies of \ce{C=O} are summarized in Table S1.

\subsection{FT-IR Spectra}

Samples for ATR-FTIR were prepared as follows.
CMPO was dissolved in IL at 0.1M and contacted with an aqueous solution of 1M NaNO$_3$ and 1mM HNO$_3$ at equal phase volumes.
This ``pre-equilibrated'' organic phase was collected and used as the extractant solution for an aqueous solution containing 20mM UO$_2$(NO$_3$)$_2$, 1M NaNO$_3$ and 1mM HNO$_3$.
The volumes of the two phases for the extraction were equal.
The phases were contacted for 30 minutes and centrifuged.
The separated organic phase with uranyl ion and the uranyl-free, pre-equilibrated organic phases were then analyzed by FT-IR.  

ATR-FTIR spectra were collected with a Nicolet iS50 ATR spectrometer equipped with a 2.8 mm round, type IIa diamond crystal, KBr beamsplitter, and a DTGS detector.
Spectra were collected with a resolution of 4 cm$^{-1}$, a zero-filling factor of 2, and 32 scans per spectrum in the wavenumber range of 4000 to 400 cm$^{-1}$.
Ambient background was subtracted from the sample spectra.
The ATR crystal was carefully cleaned with methanol between the samples.

\subsection{MD Simulations}

The \underline{G}eneral \underline{A}MBER \underline{F}orce \underline{F}ield (GAFF2) \cite{wang2004development,he2020fast} is used to describe intra- and inter-molecular interactions for CMPO, TBP, and \ce{[EMIM][Tf_{2}N]}.
Atomic partial charges for these molecules are derived from Austin Model 1 with bond charge corrections (AM1-BCC) method. \cite{jakalian2000fast,jakalian2002fast}
The parameter set for dodecane is taken from an optimized GAFF model to prevent freezing at room temperature. \cite{vo2015computational}
The SPC/E water model \cite{berendsen1987missing} is used to describe water.
\cite{izadi2016accuracy}
A uranyl model by \citeauthor{kerisit2013structure} has shown structural, kinetic, and thermodynamic properties that agree with experimental data, so we use this model for the uranyl ion in the alchemical simulations.

Geometric mixing rules are used for all cross-term Lennard-Jones interactions.
An energy cutoff of 12 \ce{\AA} is set for nonbonded Lennard-Jones and electrostatic interactions.
Long-range electrostatic interactions are accounted for by a particle-particle particle-mesh solver. \cite{hockney2021computer}
The leap-frog stochastic dynamics integrator at 350 K along with a timestep of 1 femtosecond is used for the thermostat \cite{goga2012efficient}; the Parrinello-Rahman barostat \cite{parrinello1981polymorphic,nose1983constant} is applied at 1 atmosphere.
All simulations are performed in the GROMACS MD software package. \cite{abraham2015gromacs}

\subsubsection{Umbrella sampling}

A \textit{cis}-CMPO is packed with 100 solvent molecules using PACKMOL \cite{martinez2009packmol}: 100 dodecane; 100 TBP; 100 \ce{[EMIM][Tf_2N]} for dry IL simulation; 70 \ce{[EMIM][Tf_2N]} and 30 water moleucles for wet IL simulation.
The umbrella sampling starts with an oxygen-oxygen distance of 2.8 \ce{\AA} and reaches 5.1 \ce{\AA}, with an increment of 0.1 \ce{\AA} for each window using a harominic bias potential of 100 \ce{kcal/mol\cdot\AA}.
In each window, a 20 nanosecond simulation is performed, whereas the oxygen-oxygen distance is sampled every 0.05 picosecond.
The first 1 nanosecond of NPT at each window is discarded from analysis to ensure the solvent structure around CMPO has reached equilibrium.
All free energy profiles are analyzed using the Weighted Histogram Analysis Method (WHAM).
Histograms are shown in Figures S2-S6.
Sample GROMACS input scripts and force field files (.itp) are provided at Zenodo: doi.org/10.5281/zenodo.8378117.

\subsubsection{Alchemical free energy estimations}

In order to isolate the impact of metal complexation on the solvent effect, we conduct alchemical transformation calculations on a uranyl nitrate ion pair bound to the CMPO molecule.
A \ce{UO_2(NO_3)^+} cation complex is annihilated from the IL phase to mimic the cation exchange mechanism in the uranyl extraction. \cite{visser2003uranyl}
The annihilation of a charged particle is a well-known size-effect problem in alchemical free energy calculation. \cite{rocklin2013calculating}
In order to minimize this size effect, simulations are performed for one CMPO in a box with 300 ion pairs of \ce{[EMIM][Tf_2N]}, which sizes up to roughly 50 \ce{\AA} on each dimension.
Such a box size is shown to minimize the size effect after application of a uniform background charge, which we implement in the long-range electrostatics. \cite{bogusz1998removal}
The free energy difference between the start and the end states is measured using the TI-CUBIC method by fitting the curve $\langle (\frac{\partial U(\lambda)}{\partial \lambda}) \rangle _{\lambda_{i}}$ vs $\lambda_{i}$ to a natural cubic spline and then analytically integrating through the cubic equation. \cite{bruckner2011efficiency}
This methodology has been implemented in the Alchemical Analysis code. \cite{klimovich2015guidelines}
$\langle (\frac{\partial U(\lambda)}{\partial \lambda}) \rangle _{\lambda_{i}}$ and $\Delta G$ in each window of perturbing $\lambda_{i}$ are shown in Figures S7-S11.

\section{Results and Discussions}

\subsection{FT-IR}

\begin{figure*}[h]
\centering
\includegraphics[width=1.0\linewidth]{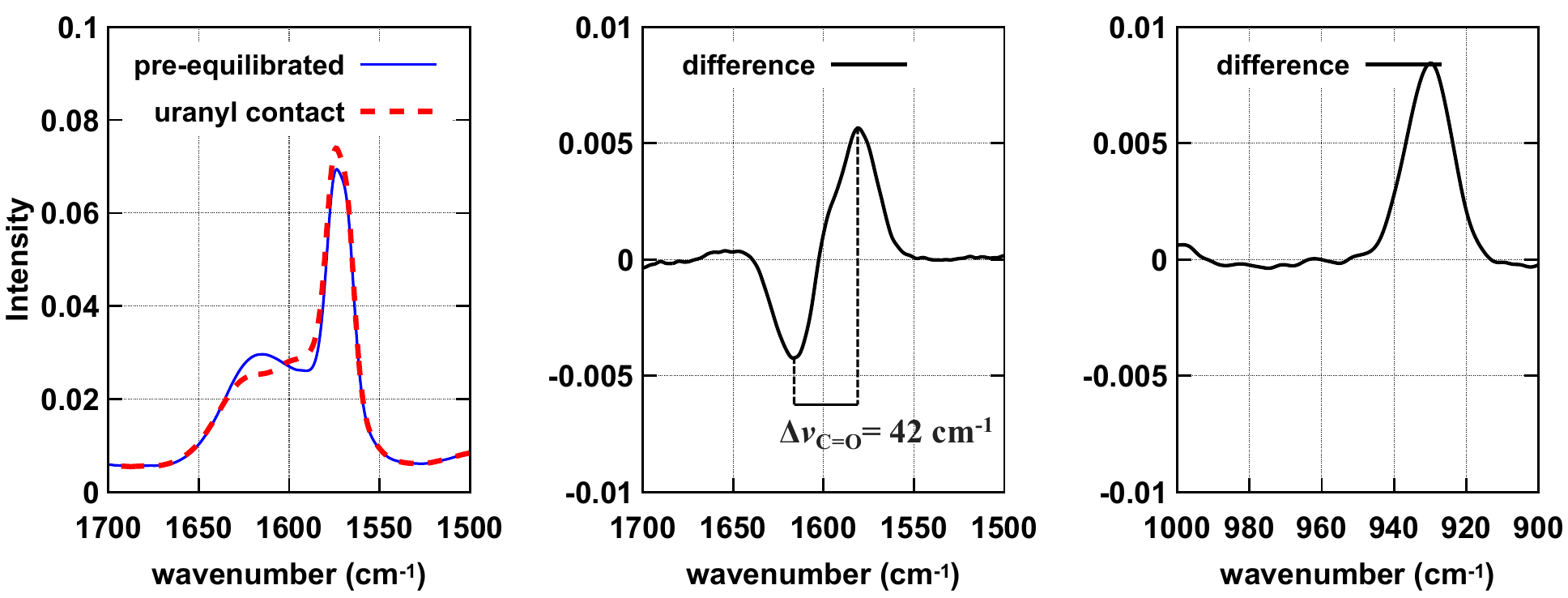}
\caption{Left panel: IR spectra of \ce{C=O} region of interest for 0.1 M CMPO in \ce{[EMIM][Tf_2N]} for the pre-equilibrated organic phase (solid line) and uranyl-contacted organic phase (dashed line). Middle panel: subtracted IR spectra of the uranyl-contacted organic phase by the pre-equibrated organic phase. Right panel: subtracted IR spectra of \ce{U=O} region of interest.}
\label{fig:ir_zoom}
\end{figure*}

First, to demonstrate that CMPO binds uranyl in a bidentate manner with the \textit{cis}-conformation even in the relatively polar IL environment, we measure FT-IR spectra after pre-equilibration with the metal-free aqueous phase and again after contact with the uranyl-containing aqueous phase.
Although \ce{P=O} stretching is in the region around 1200 cm$^{-1}$, which is difficult to distinguish from features associated with pure IL \cite{hofft2008investigations}, EXFAS studies have clearly shown \ce{P=O} binds strongly with uranyl. \cite{visser2003uranyl}
So, here we will only focus on the \ce{C=O} region: if \ce{C=O} binds with uranyl, we will have strong evidence that CMPO coordinates the uranyl in a bidentate manner.
\cref{fig:ir_zoom} shows the region of 1500--1700 cm$^{-1}$, which includes the \ce{C=O} stretching region of interest, with the difference between the IR spectra with and without uranyl shown in the right panel.
(FTIR spectra for the entire wavenumber range are shown in Figure S1.)

Although \ce{C=O} stretching is also overlapped by the strong signal of IL imidazolium \ce{C=C} stretching \cite{hofft2008investigations}, the difference in IR after uranyl extraction can provide direct evidence of the interaction between the \ce{C=O} group and the uranyl cation.
As shown in \cref{fig:ir_zoom}, the \ce{C=O} stretching mode shifts upon uptake of uranyl into the organic phase, with $\Delta\nu_{\ce{C=O}} = 0.42$ cm$^{-1}$.
This is consistent with the shift reported in the literature for the extraction of CMPO of other metal types that is attributed to bindentate complexation. \cite{tkac2012study}
The peak shift in the \ce{C=O} vibrational mode also corresponds directly to our DFT-calculated peak shift upon uranyl binding (see Table S1).
To ensure that the magnitude of the shifted peaks is not small compared to the amount of uranyl in the organic phase, we also show the change in IR spectra after extraction of uranyl in the \ce{U=O} region (900--1000 cm$^{-1}$) in \cref{fig:ir_zoom}.
This peak associated with the uranyl stretching mode is similar in intensity to the shifted \ce{C=O} peak.
Although these intensities cannot be directly and quantitatively compared, this still indicates that the approximate amount of shifted \ce{C=O} bonds is of the same order as the total amount of extracted uranyl molecules.
Overall, this suggests that CMPO forms bidentate-extracted complexes with uranyl.

\subsection{Free energy calculations and reaction coordinate}

Previously, we applied metadynamics to sample the whole free energy landscape of extractant conformation. \cite{wang2023using}
From there, we have seen significant free energy penalties for the bound-state \textit{cis} conformation in dodecane (up to 30 kJ/mol).
This large difference in energies between conformations and high energetic barriers between conformations necessitate advanced sampling techniques, such as umbrella sampling, \cite{torrie1977nonphysical} to adequately sample the entire conformational space.
Compared with the metadynamics technique in our previous study, umbrella sampling, with harmonic bias potentials, can better control collective variables along a reaction coordinate.
However, mapping this entire 2D Ramachandran plot using umbrella samling is not computationally feasible, so we instead construct a 1D reaction coordinate that differentiates the three molecular conformations of interest: the phosphine oxide to carbonyl oxygen distance.
This distance is a meaningful reaction coordinate because it readily distinguishes \textit{cis}, \textit{gauche} and \textit{trans} conformations.
By this construction of the reaction coordinate, we can reduce the 2D Ramachandran plot to a 1D map.
By performing umbrella sampling along this reaction coordinate, we can calculate the free energy differences between each conformation from MD simulations using explicitly modeled solvent.

\subsection{Gas phase, dodecane and TBP solvents}

\begin{figure}[h]
    \centering
    \includegraphics[width=1.0\linewidth]{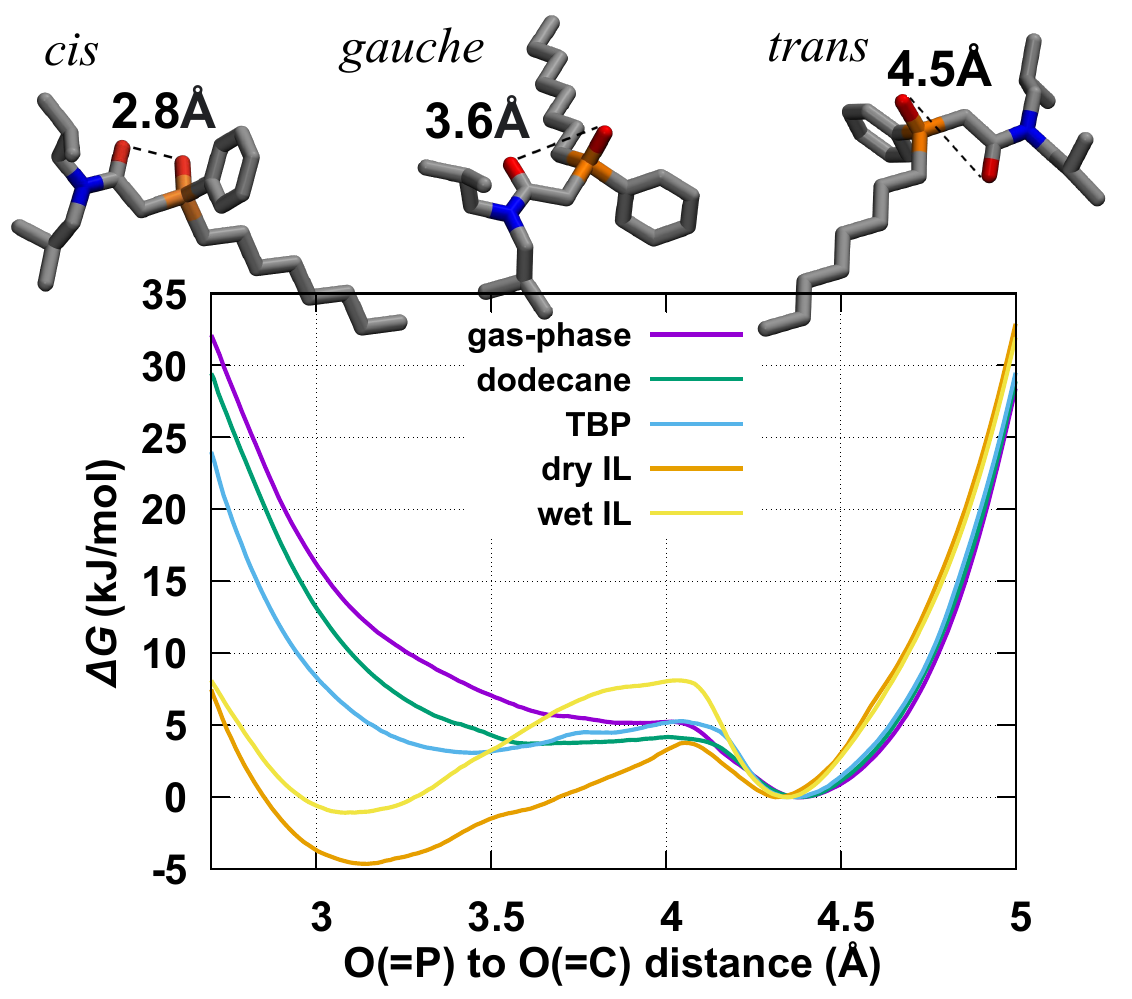}
	\caption{Free energy profiles in each solvent are plotted along the oxygen-oxygen distance reaction coordinate, with distances corresponding to each of the CMPO conformations highlighted.}
	\label{free_energy}
\end{figure}

The free energy profile in the gas phase provides information on the intrinsic energetics of the molecule in the absence of any solvent interactions.
This free energy profile along the reaction coordinate is shown in \cref{free_energy} (purple line).
The \textit{cis}-CMPO conformation corresponds to the left end of each profile at 2.8 \ce{\AA}.
The free energy levels off around 3.8 \ce{\AA}, corresponding to the \textit{gauche} conformation, while the global free energy minimum at the \textit{trans} conformation is around 4.4 \ce{\AA}.
At higher distances, the free energy continues to increase as the molecule is distorted.
Since the umbrella sampling provides the relative free energy rather than the absolute one, we anchor all profiles at $\Delta G _{trans}$ = 0 kJ/mol for cross-comparisons.
Compared to the gas phase, the dodecane solvent phase (green line) has a similar profile, with a primary difference of the dodecane profile being slightly lower energies of the \textit{gauche} and \textit{cis} regions compared to the gas phase.
Overall, both in the gas phase and in the dodecane, CMPO shows a significant free energy penalty for the \textit{cis} conformation relative to \textit{trans} (roughly \ce{\Delta G _{\textit{cis-trans}}}=30 kJ/mol).
Such a large energy penalty works against the formation of metal-extractant complexes where the free, unbound \textit{trans-}CMPO rearranges into the \textit{cis} conformation to bind the metal.

\begin{figure}[h]
    \centering
    \includegraphics[width=1.0\linewidth]{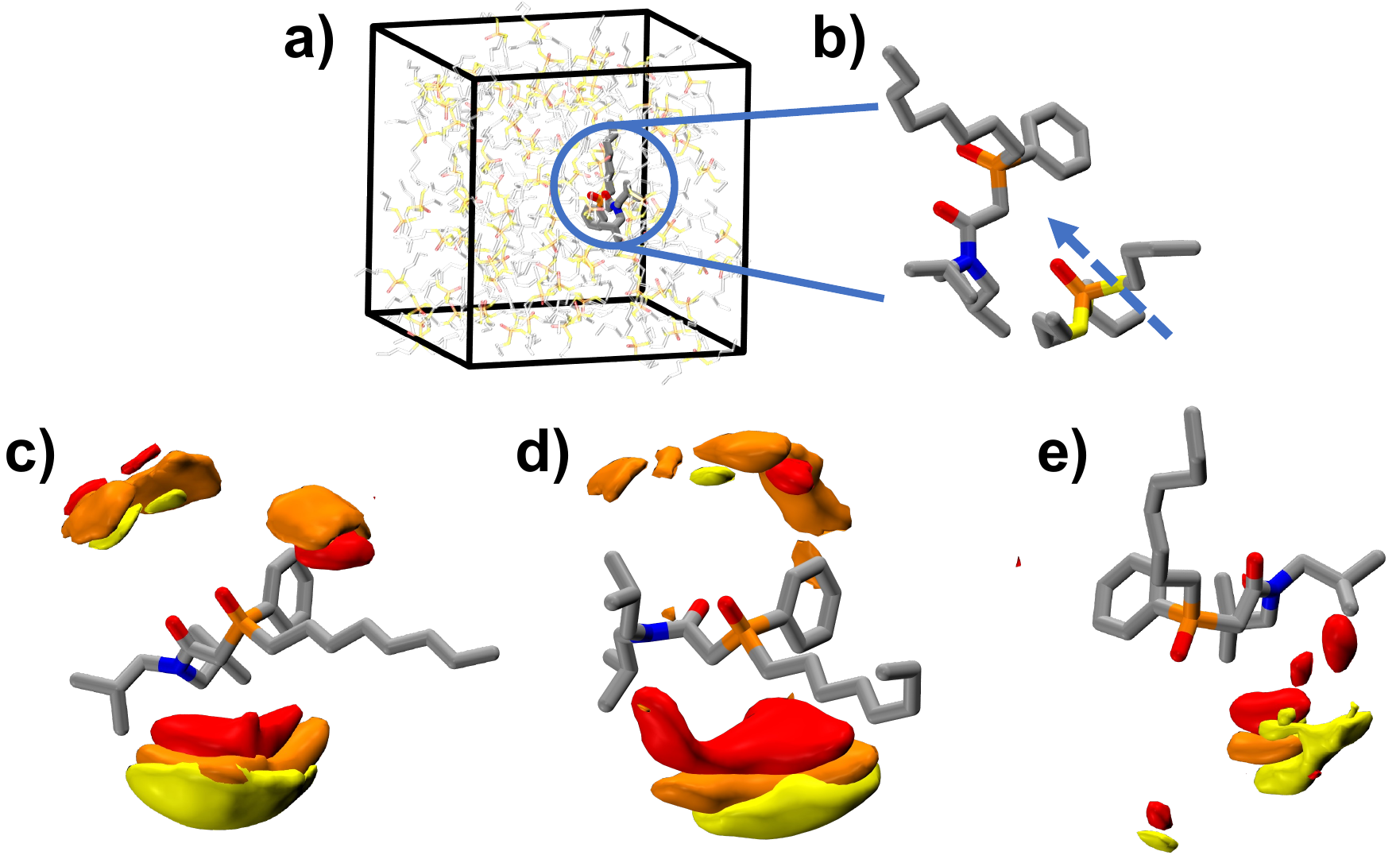}
	\caption{CMPO in TBP. a) shows the overall simulation box; b) shows a zoom-in inset of TBP alignment with \textit{cis}-CMPO, and the dashed arrow represents the overall dipole orientation of the TBP molecule; c)-e) show the spatial distribution functions (SDF) of TBP double-bonded $sp^2$ oxygen (red), phosphorus (orange) and single-bonded $sp^3$ oxygen (yellow) centered by \textit{cis}, \textit{gauche} and \textit{trans} CMPO respectively.}
	\label{tbp}
\end{figure}

While dodecane might be expected to act similarly to the gas phase, TBP is a partially polar solvent and may affect the conformational energetics of CMPO.
A snapshot of CMPO in TBP is shown in \cref{tbp}a.
For TBP, in \cref{free_energy} (blue line) we find that the free energy profile has been comparatively reduced, with \ce{\Delta G _{\textit{cis-trans}}} and \ce{\Delta G _{\textit{gauche-trans}}} decreasing by 8 kJ/mol and 10 kJ/mol, respectively.
The difference between \textit{trans} and \textit{cis} states is roughly twice as large between TBP and dodecane than dodecane and the gas phase.
A shallow local minimum is also observed around 3.4 \ce{\AA}, indicating that the \textit{gauche} conformation is stabilized by the TBP solvent.
In contrast to dodecane, TBP is partially polar and, as we find, can partially stabilize \textit{cis} and \textit{gauche} conformations through dipole alignment with the CMPO head group dipole; such solvent arrangement is shown in \cref{tbp}b.
To quantify this effect, we calculate spatial distribution functions (SDFs) to show the ensemble average alignment of TBP with CMPO for each conformation.
SDFs, often used to understand solvation in molecular simulations, show the spatial distribution of a particular atom or molecule of interest surrounding a reference atom or molecule.
\cref{tbp}c-e show the SDFs of \textit{cis}, \textit{gauche} and \textit{trans} CMPO, respectively.
For the \textit{cis} and \textit{gauche} conformations, it is clear that TBP's $sp^2$ oxygen, phosphorus, and $sp^3$ oxygen have structured distributions above and below the CMPO head group, indicating solvent TBP molecules align with the CMPO dipole. 
The \textit{trans}-CMPO has a weaker intrinsic dipole because the polar \ce{O=P} and \ce{O=C} bonds point in opposite directions.
This prevents solvent organization around the CMPO, as evidenced by the relative lack of structure in the SDF shown in \cref{tbp}e.
Overall, these results show how organization of the partially polar TBP solvent molecules around the CMPO stabilizes the more polar \textit{cis}, \textit{gauche}-CMPO conformations.

\subsection{IL solvents}

\begin{figure}[h]
    \centering
    \includegraphics[width=1.0\linewidth]{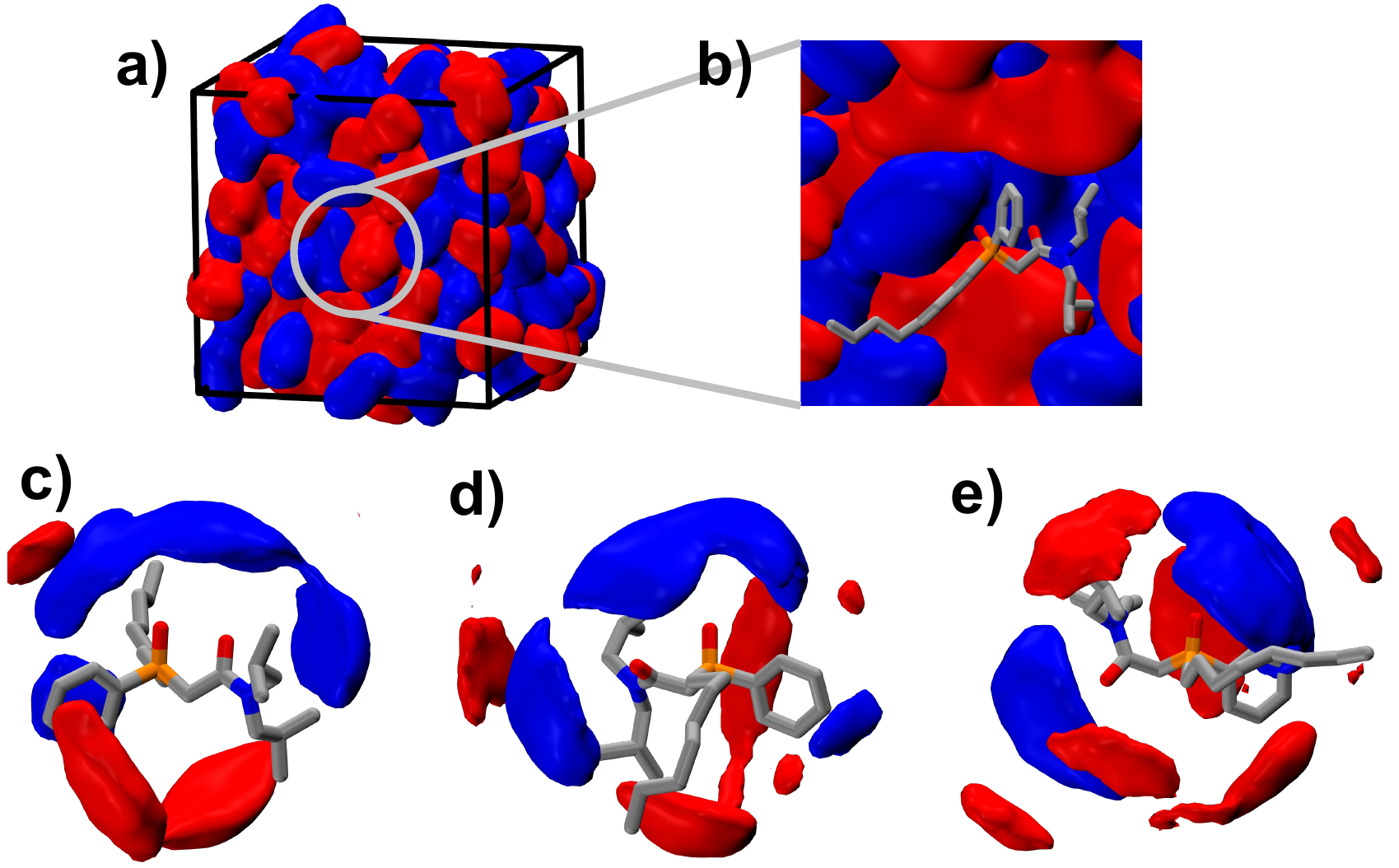}
	\caption{CMPO in dry \ce{[EMIM][Tf_2N]}. a) shows the cationic (blue) and anionic (red) domains in the simulation box; b) shows a zoom-in inset of \textit{cis}-CMPO surrounded by cationic/anionic domains; c)-e) show the spatial distribution functions (SDF) of cationic \ce{EMIM} (blue) and anionic \ce{Tf_2N} (red) centered by \textit{cis}, \textit{gauche} and \textit{trans} CMPO respectively.}
	\label{dry_IL}
\end{figure}

Next, we consider IL solvents, which are structurally unique compared to traditional molecular solvents.
In this case, the right panel in \cref{free_energy} shows the \textit{cis} and \textit{gauche} free energies are dramatically reduced.
Instead of the global minimum at the \textit{trans}, the \textit{gauche} becomes the most stable conformation.
The \ce{\Delta G _{\textit{gauche-trans}}} is  --5 kJ/mol---a reduction of 8 kJ/mol compared to dodecane; the \ce{\Delta G _{\textit{cis-trans}}} is lowered from 30 kJ/mol in dodecane to 7 kJ/mol in the dry IL.
We attribute this significant stabilization of the \textit{cis} and \textit{gauche} conformations to the alignment of the CMPO dipole with the cation/anion nanodomains characteristic of ILs. \cite{hayes2015structure,kashyap2013structure}
\cref{dry_IL}a-b shows a representative snapshot in which CMPO is surrounded by cationic/anionic domains.
More importantly, the SDFs in \cref{dry_IL} show that, for both \ce{O=P} and \ce{O=C} groups, strong negative partial charges of the $sp^2$ oxygen atoms lead to cationic \ce{[EMIM]} domains surrounding those oxygen atoms.
Conversely, the other partially positive end of the \ce{O=P} and \ce{O=C} dipoles associate with the anionic \ce{[Tf_2N]} domains.
The cationic and anionic layers therefore induce dipole alignment of the embedded extractant molecule, favoring the \textit{cis} and \textit{gauche} conformations.
Considering that multidentate extractants, such as CMPO, may need to overcome a high energy barrier to reach their metal-binding conformation, such significant stabilization induced by the solvent could potentially improve separations performance.
This observation implies new opportunities for further control of separation energetics, such as the customization of nanodomains through IL design.

\begin{figure}[h]
    \centering
    \includegraphics[width=1.0\linewidth]{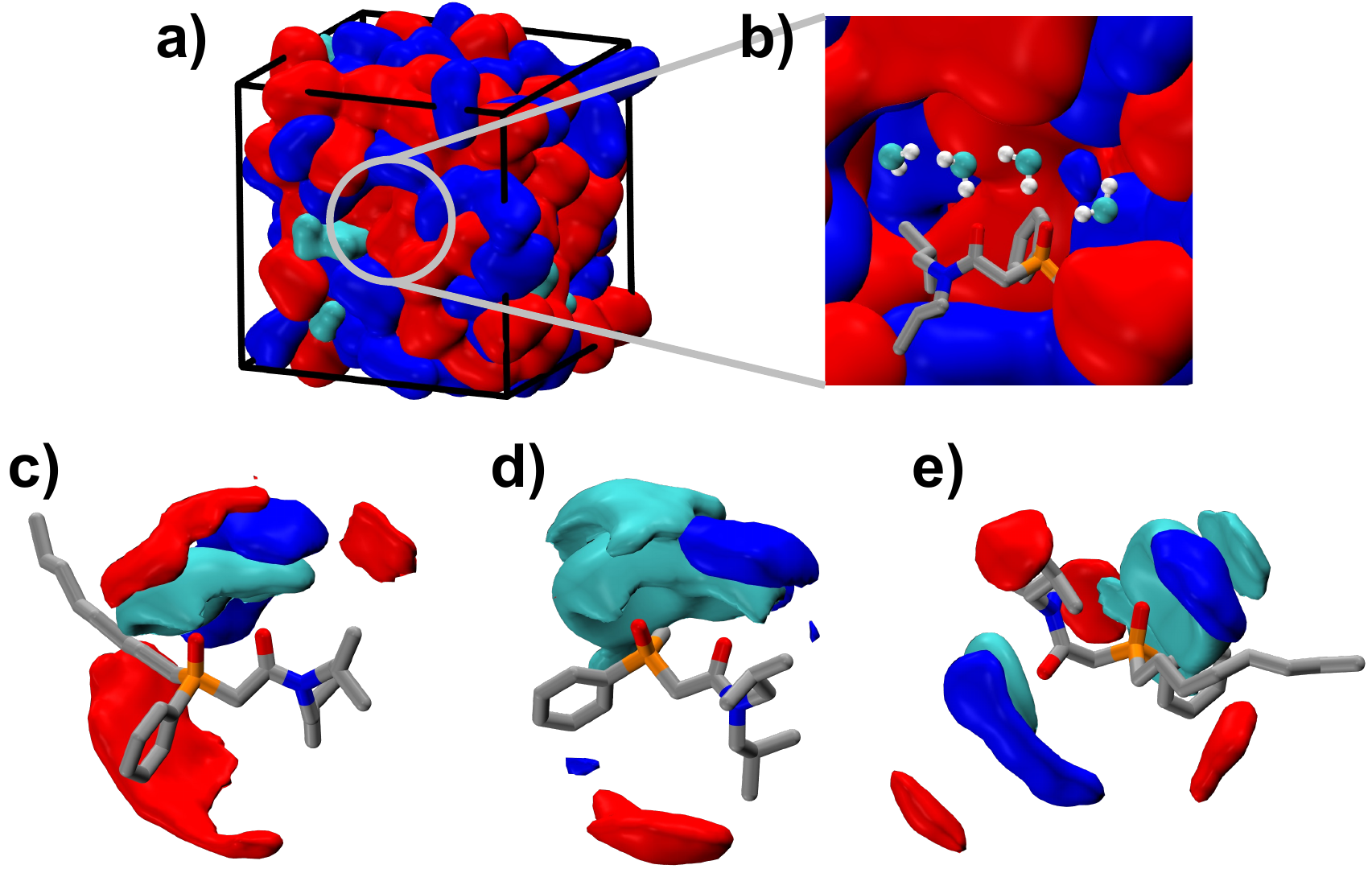}
	\caption{CMPO in wet \ce{[EMIM][Tf_2N]}. a) shows the cationic (blue), anionic (red), and water (cyan) domains in the simulation box; b) shows a zoom-in inset of \textit{cis}-CMPO surrounded by cationic/anionic domains while forming hydrogen bond with water molecules; c)-e) show the spatial distribution functions (SDF) of cationic \ce{EMIM} (blue), anionic \ce{Tf_2N} (red) and water (cyan) centered by \textit{cis}, \textit{gauche} and \textit{trans} CMPO respectively.}
	\label{wet_IL}
\end{figure}

The solubility of water in ILs can be significant and, in LLE applications where the IL is in direct contact with an aqueous phase, nonnegligible amounts of water transfer to the IL phase.
For example, the saturated concentration of water in \ce{[EMIM][Tf_2N]} is 0.2982 mole fraction, or 2.89\% volume fraction (vol\%). \cite{freire2008mutual}
Therefore, to account for the realistic solution environment encountered in LLE, we also consider the water-saturated IL.
Despite hydrogen bonding between water and CMPO's polar \ce{O=P} and \ce{O=C} sites, the relative free energies of the CMPO conformations are nearly the same as the dry IL (\cref{free_energy}).
A zoom-in inset in \cref{wet_IL}b shows two water molecules are forming hydrogen bonds with the $sp^2$ oxygen atoms on CMPO.
SDFs in \cref{wet_IL}c-e further illustrate such molecular events: we can see that water distributes (cyan) around the two $sp^2$ oxygen atoms for all three \textit{cis}, \textit{gauche} and \textit{trans} CMPO conformations.
This finding suggests that any impact of water hydrogen bonding favoring \textit{cis} or \textit{gauche} conformations in the CMPO is either not significant or roughly cancelled out by the water interfering with the dipole alignment induced by the IL nanodomains.

\subsection{Alchemical free energy calculations with uranyl ion}

The above analysis considers the impact of the solvent in the absence of metal ions.
In an actual extraction process, the metal ion will participate in the coordination with the extractant molecule.
Such participation of the metal ion will potentially diminish the solvent preorganization effect onto the extrantant energetics, especially when one considers that part of the extractant will be binding to the metal ion, potentially suppressing interactions with the surrounding solvent.
To directly test the solvent effect with the presence of the metal ion, we perform a sequence of alchemical transformation calculations.
We begin alchemical simulations with an assembled \ce{UO_2(NO_3)(CMPO)} complex. Realizing that in a real extraction system, the IL phase would contain a nonnegligible amount of water after contact with the aqueous phase, which could compete with nitrate for the coordination to uranyl, in our alchemical calculations we choose the dry IL system rather than the wet one. Such a choice will certainly have an impact on the metal coordination environment. However, in this work, we want to focus on how the IL nanodomains interact with the extractant rather than the impact of water molecules. In our future work, we plan to add water to the simulations and test its contribution to the extraction energetics.
By calculating the solvation free energy of a uranyl nitrate bound to CMPO with and without a bias potential on the extractant conformation, we can ascertain the free energy cost of constraining the extractant to the binding conformation and compare it to the same cost in the absence of the metal.
The difference, then, reveals how much the cation has diminished the effect of the solvent.
By comparing this to the free energy profiles above obtained by umbrella sampling, we can identify which fraction of the solvent reorganization free energy resulting from the solvent persists in the presence of the metal.

\begin{figure}[H]
\centering
\includegraphics[width=1.0\linewidth]{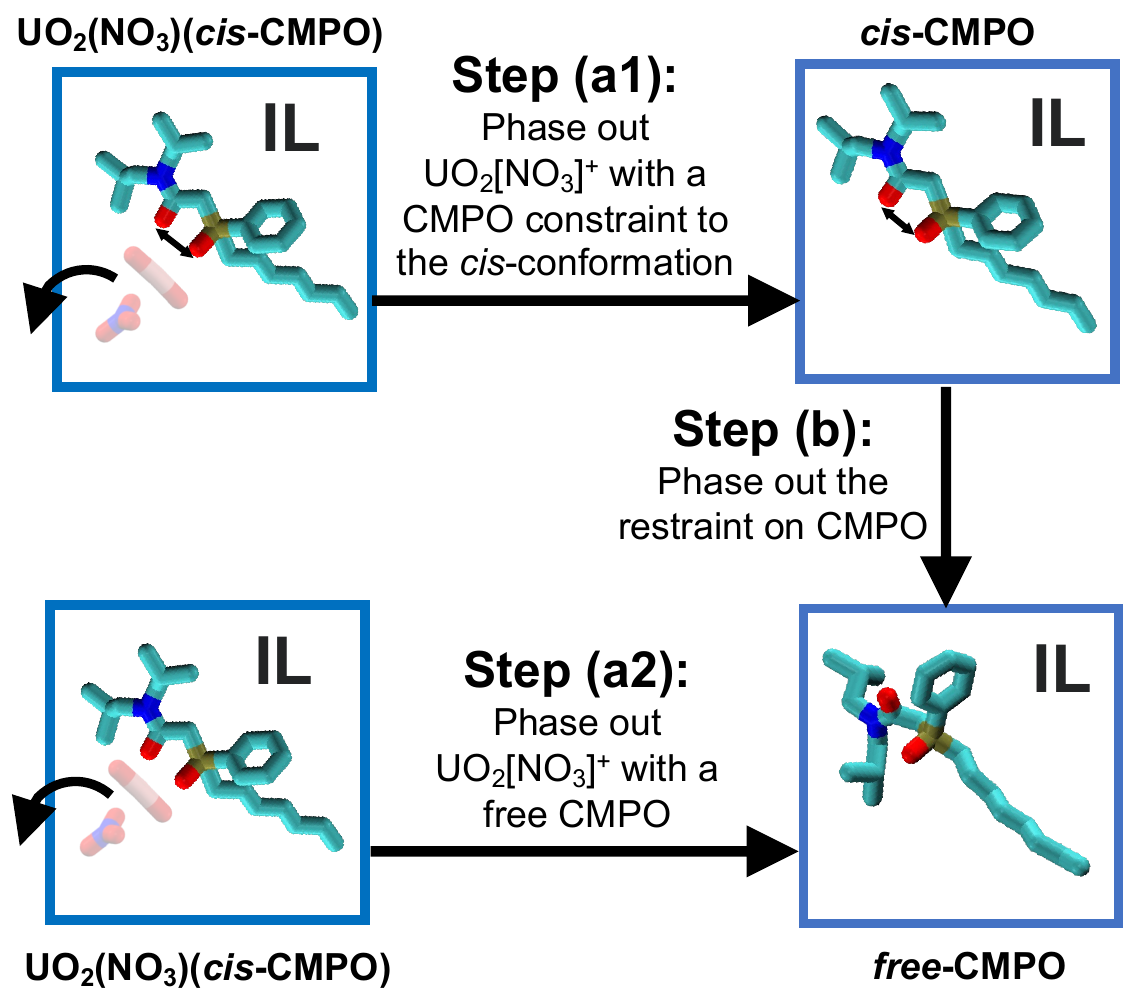}
\caption{Schematic flow chart of alchemical transformations calculated. Step (a1): Annihilating \ce{UO_2(NO_3)^+} with a restraint placed to keep CMPO stay at \textit{cis}-conformation. Step (a2): Annihilating \ce{UO_2(NO_3)^+} without any restraint on CMPO. Step (b): Annihilating the restraint on CMPO to let it relax from \textit{cis}-conformation to free.}
\label{fig:alchemy_flow}
\end{figure}

A schematic flow chart of how alchemical transformations are carried out in this work is shown in \cref{fig:alchemy_flow}.
Details associated with each alchemical transformation are described below.

\begin{itemize}
    \item [(a1)] \textbf{The annihilation \ce{UO_2(NO_3)^+} with a \textit{cis}-CMPO:} One \ce{UO_2^{2+}} and one \ce{NO_3^{-}} are contained in the simulation box. In the initial configuration of the metal complexant \ce{UO_2(NO_3)(CMPO)}, \ce{CMPO} and \ce{NO_3^{-}} chelate with uranyl in a bidentate fashion. The \ce{UO_2^{2+}} and \ce{NO_3^{-}} are decoupled from the simulation box by mutating $\lambda_{vdw}$ and $\lambda_{coul}$ in the soft-core version of the 12-6-1 potential. $\lambda_{coul}$ is decoupled first followed by $\lambda_{vdw}$, in order to prevent the ``end-point catastrophes''. 12-point scheme is chosen for $\lambda_{coul}$ (0.00922, 0.04794, 0.11505, 0.20634, 0.31608, 0.43738, 0.56262, 0.68392, 0.79366, 0.88495, 0.95206, 0.99078) to minimize quadrature errors. For $\lambda_{vdw}$, an increment of 0.1 is chosen between each window. To keep CMPO at \textit{cis}-comformation, a harmonic restraint of 100 \ce{kcal/mol\cdot\AA} is placed between phosphine oxide and carbonyl oxygen to maintain the distance at 2.8 \ce{\AA}.

    The rationale behind the annihilation of a \ce{UO_2(NO_3)^+} ion rather than a neutral ion pair of \ce{UO_2(NO_3)_2} is due to the cation-exchange mechanism of interest in this IL system. For the divalent uranium separation using CMPO as extractant and \ce{[EMIM][TF_2N]} as diluent, the cation-exchange mechanism is that a uranyl ion forms a complex with one nitrate and one CMPO extractant -- \ce{UO_2(NO_3)(CMPO)^+}. Then, the extraction is achieved by swapping a \ce{[EMIM]^+} from the IL phase into the aqueous phase. This mechanism is well established and has been extensively tested using slope analysis and EXAFS by \citeauthor{visser2003uranyl} \cite{visser2003uranyl}.

    \label{alchemy1}
    \item [(a2)] \textbf{The annihilation of \ce{UO_2(NO_3)^+} with a free CMPO:} This alchemical transformation is the same as in the previous step, except for the restraint on the CMPO. We do not apply any restraint on the CMPO molecules, thus, it can freely move in solution, while the \ce{UO_2^{2+}} and \ce{NO_3^{-}} are phased out.
    \label{alchemy2}
    \item [(b)] \textbf{The annihilation of the restraint on \textit{cis}-CMPO:} In this step, in order to test how much energy can be recovered from relaxing the CMPO conformation, we phase out the restraint from the end state in step 1. The restraint is decoupled by linearly scaling down the restraint to 0 \ce{kcal/mol\cdot\AA} with 21 windows.
    \label{alchemy3}
\end{itemize}

\begin{figure}[H]
    \centering
    \includegraphics[width=1.0\linewidth]{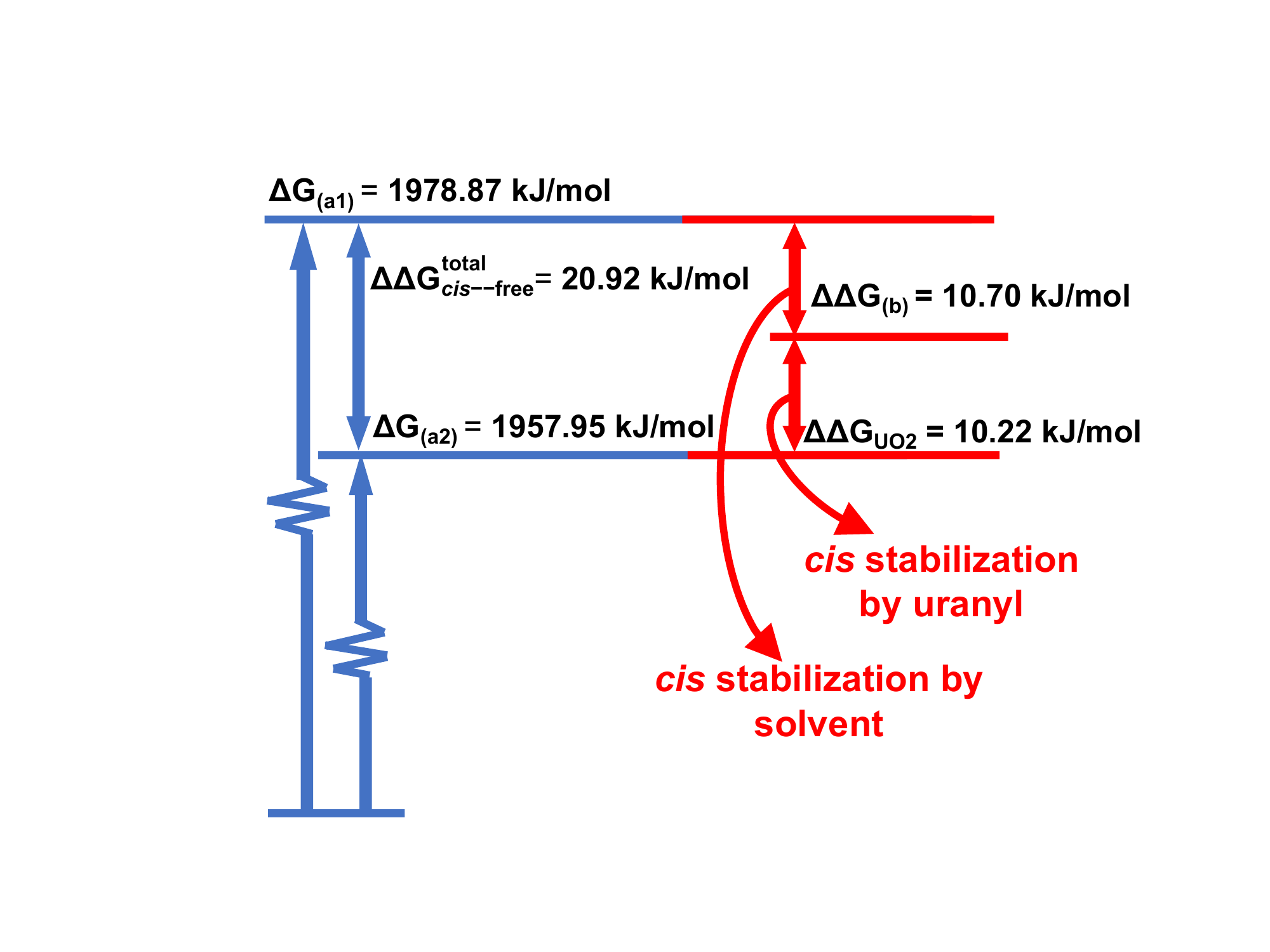}
	\caption{Free energies of each step in the alchemcial transformation calculations. Steps (a1) and (a2) involving annihilating the uranyl ion have been colored by blues, while step (b) involving the extractant pre-organization without the uranyl ion is colored by red.}
	\label{fig:alchemical}
\end{figure}

From the alchemical transformations outlined above, we define the relevant free energies given in \cref{eq:delta1,eq:delta2}.
\begin{equation}
    \Delta\Delta G^{\text{total}}_{cis-\text{free}} = \Delta G_{\text{(a1)}} - \Delta G_{\text{(a2)}}
    \label{eq:delta1}
\end{equation}
\begin{equation}
    \Delta\Delta G^{}_{\text{UO}_2} = \Delta\Delta G^{\text{total}}_{cis-\text{free}} - \Delta\Delta G_{\text{(b)}}
    \label{eq:delta2}
\end{equation}
$\Delta\Delta G^{\text{total}}_{cis-\text{free}}$ is the free energy difference of solvating the uranyl between the \textit{cis}-constrained and unconstrained CMPO molecule.
$\Delta\Delta G^{}_{\text{UO}_2}$ is the contribution to $\Delta\Delta G^{\text{total}}_{cis-\text{free}}$ that results from the presence of the uranyl, i.e., the contribution that does \textit{not} come from just the solvent in the absence of the uranyl.
\ce{\Delta G_{(a1)}}, \ce{\Delta G_{(a2)}}, and \ce{\Delta\Delta G_{(b)}} are from the alchemical free energy calculations in steps (a1), (a2), and (b) as we describe above.
We visually show the relationship of these free energy terms in \cref{fig:alchemical}.
From the same initial configuration, steps (a1) (annihilation of uranyl with CMPO restrained) and (a2) (annihilation of uranyl with CMPO free) result in different end states --- \textit{cis} and free CMPO in IL, respectively.
These two alchemical transformations have also resulted in a notable difference in free energies, with a \ce{\Delta\Delta G^{total}_{\textit{cis}-free}} of 20.92 kJ/mol.
Noting that annihilation of the uranyl nitrate with CMPO restrained to the \textit{cis} conformation requires a higher energy (\ce{\Delta G_{(a2)}} $<$ \ce{\Delta G_{(a1)}}), this observation indicates that the uranyl ion is better solvated by \textit{cis}-CMPO.
To prove that these numbers are not affected by the error when the curves of $\langle (\frac{\partial U(\lambda)}{\partial \lambda}) \rangle _{\lambda_{i}}$ vs. $\lambda_{i}$ are fitted to cubic splines, in addition to TI-CUBIC, we have carried out extra free energy estimations using BAR \cite{bennett1976efficient} and MBAR \cite{shirts2008statistically}.
The comparisons have been shown in Figure S8, S10; differences between these methodologies are minimal.

The difference in free energy between two end states, \ce{\Delta\Delta G^{total}_{\textit{cis}-free}}, includes the stabilization effects of the uranyl metal ion and the IL cationic/anionic nanodomain.
Although these two end states do not contain the uranyl metal ion, its influence comes from the steps (a1) and (a2) during the annihilation of metal ion.
By then phasing out the restraint that keeps the CMPO at the \textit{cis} conformation, 10.70 kJ/mol of free energy (\ce{\Delta\Delta G_{(b)}}) is recovered in step (b).
It is worth mentioning that, in this case, TI-CUBIC could give sizable uncertainty when fitting the curve of $\langle (\frac{\partial U(\lambda)}{\partial \lambda}) \rangle _{\lambda_{i}}$ vs. $\lambda_{i}$ using a cubic spline, which is shown in Figure S11.
This uncertainty becomes more relevant in the last window, where an abrupt decrease of $\langle (\frac{\partial U(\lambda)}{\partial \lambda}) \rangle _{\lambda_{i}}$ vs. $\lambda_{i}$ occurs.
As a result, fitting $\langle (\frac{\partial U(\lambda)}{\partial \lambda}) \rangle _{\lambda_{i}}$ in this region using a cubic spline could potentially lead to error in the estimated free energy.
Therefore, to avoid this issue, we use MBAR, which has shown to provide minimal variance by making use of equilibrium data collected from multiple states, \cite{shirts2008statistically} to estimate the free energy for this step, finding a value of \ce{\Delta\Delta G_{(b)}} = 10.70 kJ/mol.
To show the reliability of MBAR over TI-CUBIC, we also show the overlap matrix in Figure 12S, which has satisfied overlapping for the last several $\lambda_{i}$s.
This \ce{\Delta\Delta G_{(b)}} from the alchemical calculations is slightly smaller than \ce{\Delta G_{\textit{cis-gauche}}} in \cref{free_energy} from umbrella sampling (showed by the orange line), which is presumably due in part to how, when the bias potential is removed from the \textit{cis}-CMPO, we are sampling an ensemble average of all lower energetic configurations for free the CMPO, rather than the lowest \textit{gauche} configuration in \cref{free_energy}. 
Although alchemical and umbrella sampling are two types of free energy methodologies based on different theories, the observation that \ce{\Delta\Delta G_{(b)}} and \ce{\Delta G_{\textit{cis-gauche}}} agree with each other in a reasonable way provides extra confidence in both sets of free energy calculations.

To verify that \ce{\Delta\Delta G_{(b)}} comes from the solvent effects, we also perform step (b) in the gas phase.
With the change in solvent background, \ce{\Delta\Delta G_{(b)}} increases to 27.68 kJ/mol, again agreeing the umbrella sampling result for \ce{\Delta G_{\textit{cis-trans}}} in \cref{free_energy} (shown by the purple line).
After cross-checking the values for \ce{\Delta\Delta G_{(b)}} from the solvent effect, we can subtract this contribution from \ce{\Delta\Delta G^{total}_{\textit{cis}-free}}, resulting in 10.22 kJ/mol (\ce{\Delta\Delta G_{UO$_2$}}) from stabilization of the \textit{cis} conformation through metal-extractant binding.
From breaking down each contribution and comparing \ce{\Delta\Delta G^{total}_{\textit{cis}-free}}, \ce{\Delta\Delta G_{(b)}} and \ce{\Delta\Delta G_{UO$_2$}}, we therefore conclude that roughly half of the stabilization of the \textit{cis} conformation from the IL solvent is retained in the presence of the extracted metal ion.
If the stabilization imparted by the anion domain is unaffected by the replacement of the IL cation domain with the metal cation upon binding, perhaps approximately half of the total energetic stabilization of the \textit{cis} conformation would be expected.

\section{Conclusions}

Reversible LLE processes are typically driven by small free energy differences in the solubility of the targeted solute between immiscible phases; however, individual steps in the metal complexation and extraction process are often characterized by relatively larger energetic contributions that mostly cancel out.
These include the electrostatic interactions between extractant and metal, as well as the assembly of multiple extractants into the final extracted complex.
In this study, we find that the choice of organic solvent induces substantial changes in the energetic cost associated with organizing the extractant into the binding conformation.
For example, the roughly 20 kJ/mol free energy difference between binding and non-binding conformations we find between dodecane and IL is large compared to the typical total extraction free energies.
Then, by directly introducing the metal ion in the simulations and calculating the free energy solvation from the alchemical transformation of annihilating the metal ion with both free and \textit{cis}-restrained CMPO, we are able to separate stabilization effects of the metal from those of the IL solvent.
These alchemical free energy calculations show that roughly half of the extractant reorganizational free energy results from just the solvent effects measured with umbrella sampling even in the presence of the metal.
The magnitude of this effect, with a calculated value of 10.79 kJ/mol, is sufficiently large to have a substantial impact on practical separations performance.
Since LLE is often driven by very small free energy around a few kJ/mol, the stabilization effect on extractant energetics induced by IL nanostructures is not negligible. This observation can potentially explain, at least in part, the order-of-magnitude increase in distribution ratio revealed by \citeauthor{visser2003room} \cite{visser2003room}.

Our molecular dynamics simulations reveal that the mechanism for stabilizing the binding conformation in the IL is alignment with the anion/cation nanodomains, which create layers of positive and negative charge that promote dipole alignment of the extractant molecule, thereby stabilizing the more strongly dipolar \textit{cis} conformation.
Overall, these results suggest that considering solvent effects on extractant conformation directly and explicitly is essential to broadly understand and manipulate the delicate balance of extraction energetics.
Such a mechanism by which the solvent induces a significant change in the extractant's conformational free energy landscape indicates broad possibilities for tailoring solvents to manipulate separation energetics.

\begin{acknowledgement}

This work was supported by U.S. Department of Energy (DOE), Office of Science, Office of Basic Energy Sciences, Chemical Sciences, Geosciences, and Biosciences Division, Separation Science Program, under Contract DE-AC02-06CH11357 to UChicago Argonne, LLC, Operator of Argonne National Laboratory.
We gratefully acknowledge the computing resources provided on Bebop, a high-performance computing cluster operated by the Laboratory Computing Resource Center at Argonne National Laboratory.
We thank Dr. Peter Tkac and Dr. M. Alex Brown for generously providing CMPO.
We thank Dr. Ilya A. Shkrob for helping with the NMR measurement.

\end{acknowledgement}

\begin{suppinfo}

IR spectra experiments; histograms from umbrella sampling along the reaction coordinate in dodecane, TBP, dry and wet IL; plots of $\langle (\frac{\partial U(\lambda)}{\partial \lambda}) \rangle _{\lambda_{i}}$ and $\Delta G$ vs $\lambda$ in the alchemical free energy calculations.

\end{suppinfo}

\bibliography{reference}

\end{document}